\documentclass[aps,prb,reprint,nofootinbib]{revtex4-1}
\pdfoutput=1
\usepackage{amsmath} 
\usepackage{amsthm} 
\usepackage{amssymb}	
\usepackage{graphicx} 
\usepackage[squaren,Gray]{SIunits}
\usepackage{subfigure}
\usepackage{multirow}
\usepackage{makecell}
\usepackage[textsize=tiny,backgroundcolor=white]{todonotes}
\usepackage{lipsum}
\usepackage{stmaryrd}
\usepackage{hyperref}
\graphicspath{ {img/} }
\makeatletter
\def\input@path{ {content/}}
\makeatother

\let\baraccent=\= 
\renewcommand{\=}[1]{\stackrel{#1}{=}} 

\theoremstyle{definition}

\theoremstyle{remark}

\begin{document}
\title{Compact Extra Dimensions in Quantum Mechanics}

\author{Nicolas Deutschmann}
\email{n.deutschmann@ipnl.in2p3.fr} 
\affiliation{Univ. Lyon, Universit\'e Claude Bernard Lyon 1, CNRS/IN2P3, UMR5822 IPNL, F-69622, Villeurbanne, France}
\affiliation{Center for Cosmology, Particle Physics and Phenomenology (CP3),
Universit\'e catholique de Louvain,
Chemin du Cyclotron 2, 1348 Louvain-La-Neuve, Belgium.}

\date{\today}

\begin{abstract}
Extra dimensions are a common topic in popular descriptions of theoretical physics with which undergraduate student most often have no contact in physics courses. This paper shows how students could be introduced to this topic by presenting an approach to two basic consequences of the presence of compact extra dimensions based on undergraduate-level physics. The insensitivity of low-energy physics to compact extra dimensions is illustrated in the context of non-relativistic quantum mechanics and the prediction of Kaluza-Klein excitations of particles is discussed in the framework of relativistic wave equations. An exercise that could be used as a follow-up to the ``particle in a box'' is proposed.
\end{abstract}

\maketitle 

\section{Introduction}
Extra dimensions are probably the best-known feature of string theory, thanks to a number of discussions in science outreach books\cite{randallbook}, shows\cite{elegantvid}, or other initiatives\cite{stringweb}.
Physics undergraduate students are generally aware of the existence of the concept, but have little to no formal exposure to it, since extra dimensions appear mostly in discussions of string theory or extensions of the standard model of particle physics, both of which are typically featured at the graduate level.
While the development of extra dimensions in modern, realistic theories do require advanced skills like quantum field theory, it is however possible to give substance to the idea using regular quantum mechanics.

This paper attempts to build a bridge between the discussion of extra dimension in popular science and their full-fledged realization in extensions of the standard model\cite{edreview} and string theory\cite{stringwitten}.
It is designed to be approachable by students that have dealt with the canonical exercise of a particle in a box\cite{boxexo}, which usually appears early in quantum mechanics courses.
In practical terms, the objective of this article is twofold:
\begin{itemize}
  \item to give a precise meaning to the popularized statement that ``small extra dimensions are not visible in low energy physics''\cite{elegantbook}, using a similar approach to the treatment in Ref.~\onlinecite{zwiebach},
  \item to explain one of the most general experimental predictions of extra dimensions.
\end{itemize}

We will however not discuss the more general context for extra-dimensional theories, the motivation for extra dimensions in particle physics \cite{UEDcontext} or other features of string theory.

This paper will first develop in Section \ref{cyl} the simplest realization of the interplay between usual (infinite) dimensions and small compact dimensions: a particle on a cylinder. Two more complete examples with a 3-dimensional space and one and two extra dimensions will then be studied in Section \ref{3D} in order to build intuition about the general case, which will be treated briefly. Section \ref{relat} will then extend the discussion -- more sketchily -- to the case of relativistic theories and derive their main prediction. Finally, the Appendix proposes an example of exercise working out the same steps as the paper, that educators could propose to students as a follow up to the particle in a box.

\section{Free particle on a cylinder}
Let us start in a simplified situation to make the physics clear.
We will consider a particle moving freely on an infinite cylinder of radius $R$: this provides a situation with one infinite direction, along the cylinder, and one compact direction, around the cylinder.
Modelling a particle on this surface is quite simple: we will consider a wavefunction $\Psi(x,y)$, on which we impose a periodicity condition $\Psi(x,y) = \Psi(x,y+2\pi R).$
The coordinate $x$ therefore spans the infinite dimension, often dubbed $z$ in cylindrical coordinates and $y$ corresponds to $R\times\theta$.

As we are interested in the movement of a free particle, the Hamiltonian operator is simply
\begin{equation}
{\cal H}_\text{cyl} = - \frac{\hbar^2}{2 m}\partial_x^2 - \frac{\hbar^2}{2 m} \partial_y^2.
\end{equation}

Changes from the usual free particle will appear from the imposition of the boundary condition, much like in the case of the particle in a box.

The form of the Hamiltonian strongly suggests the Ansatz $\Psi_\lambda = \exp(i k_x x + i k_y y).$
Such a function is indeed an eigenfunction with eigenvalue
\begin{equation}
\lambda\left(k_x,k_y\right) = \frac{\hbar^2}{2m}\left(k_x^2+k_y^2\right).
\end{equation}
Only a subset of these functions, however, verify the periodicity condition: if $\Psi_\lambda\left(x,y\right) = \Psi_\lambda\left(x,y+2 \pi R\right)$, then $k_y$ must be a multiple of $1/R$.

The energy spectrum is therefore parametrized by two quantum numbers:

\begin{equation}
  E_n\left(k_x\right) = \frac{\hbar^2}{2m}(k_x^2+\frac{n^2}{R^2});\text{  } k_x \in \mathbb{R},\, n\in\mathbb{N}.
\end{equation}

\begin{figure}[!h]
  \centering
  \includegraphics[width=0.4\textwidth]{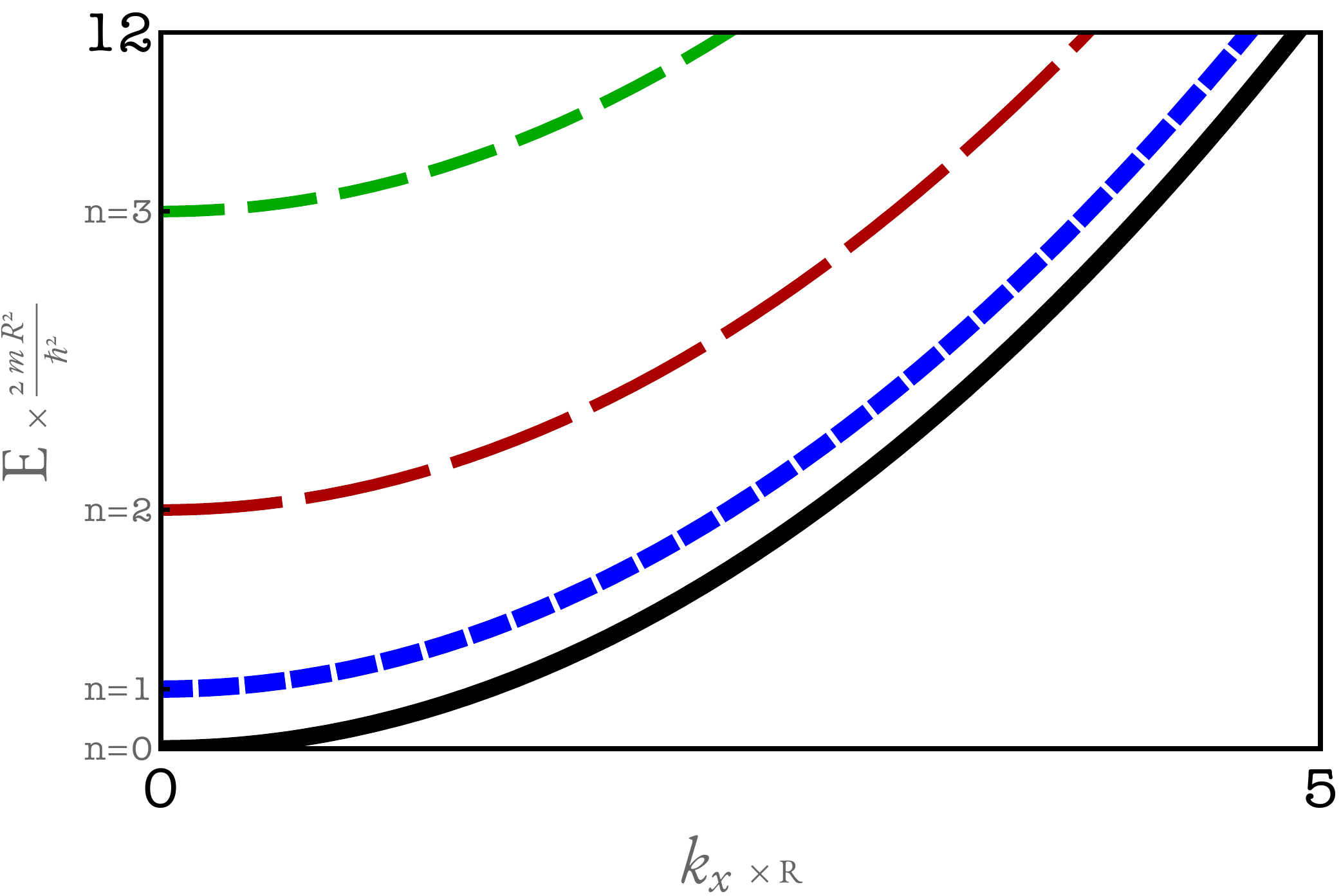}
  \caption{Plot of the eigenenergies for a free particle on a cylinder.}
  \label{specplot}
\end{figure}

Which we can graphically represent in the plane $(E_n,k_x)$ for different values of $n$ as in Fig. \ref{specplot}

The spectrum is composed of continuous bands for fixed values of $n$, which each span from a minimum value ${\cal E}_n$ to infinity, with

\begin{equation}
{\cal E}_n=\frac{n^2\hbar^2}{2mR^2}.
\end{equation}

\begin{figure}[!h]
  \centering
  \includegraphics[width=0.4\textwidth]{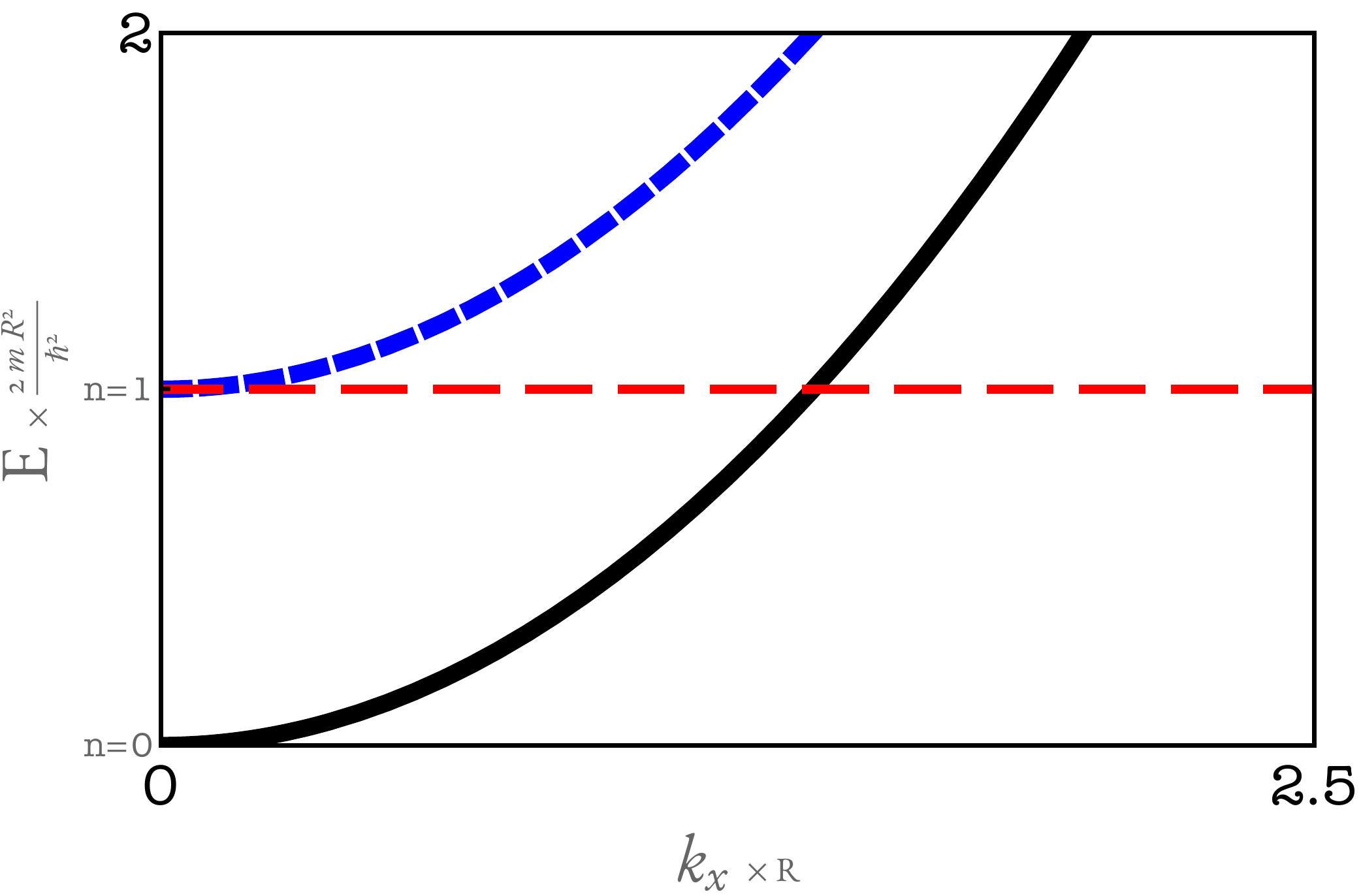}
  \caption{Low-lying eigenenergies for a free particle on a cylinder. Below the horizontal dashed line at ${\cal E}_1$, the spectrum is identical to that of a free 1D particle.}
  \label{specplotzoom}
\end{figure}

Let us focus on the low-energy part of this spectrum. In particular, at energies below ${\cal E}_1$, a single continuous band, $n=0$, is accessible, as shown in Figure \ref{specplotzoom}. As a consequence, there is a one-to-one relationship between the energy of the system below ${\cal E}_1$ and the momentum along the cylinder $p_x=\hbar\,k_x$:

\begin{equation}
E\left(k_x\right) = \frac{\left(\hbar k_x\right)^2}{2 m} = \frac{p_x^2}{2 m}.
\end{equation}

Which is exactly the energy-momentum relation of a one-dimensional free particle. In fact, notice that the energy is completely independent of the radius, which stays valid as long as the energy is below ${\cal E}_1$, or equivalently, as long as $k_x^{-1}>R$: low-energy, long-wavelength physics is independent of short-distance phenomena. Yet another manifestation of this property is visible at the level of the wavefunction:
$\Psi_{0,k_x}(x,y) = \exp\left(i k_x x\right),$
which is independent of $y$. Hence, the particle is moving along the cylinder and not around it:

\begin{equation}
  \begin{array}{l}
\hat p_x \Psi_{0,k_x}= (-i \hbar \partial_x) \Psi_{0,k_x} = \hbar k_x,\\
\hat p_y \Psi_{0,k_x}= (-i \hbar \partial_y) \Psi_{0,k_x} = 0.
\end{array}
\end{equation}

This captures the essence of why compact dimensions are invisible at low energies. In order to ground this statement more firmly, we will now turn to more realistic cases in which we will see that it is true in general for energies smaller than ${\cal O}(\hbar^2/(2mL^2))$, where $L$ is the typical length scale of the extra dimensions.

\label{cyl}
\section{Free particle in a spacetime with extra-dimensions}
\label{3D}
\subsection{One extra dimension}
It is now easy to generalize our conclusions to a more realistic situation in which there are 3 usual dimensions with coordinate $\vec x = (x,y,z)$ and $d$ compact extra dimensions with coordinates $(\zeta_1,\dots,\zeta_d)$.
Let us first start again with a circular extra dimension, \textit{i.e.} with a single extra coordinate $\zeta_1$ which is $2 \pi R$-periodic.
The $3+1D$ Hamiltonian is:

\begin{equation}
{\cal H}_\text{3+1} = - \frac{\hbar^2}{2 m}\partial_x^2 - \frac{\hbar^2}{2 m}\partial_y^2 - \frac{\hbar^2}{2 m}\partial_z^2 - \frac{\hbar^2}{2 m} \partial_{\zeta_1}^2.
\end{equation}

It is clear that a plane-wave Ansatz is once more the right form for the eigenfunctions.
After imposing the periodicity condition on the circle momentum, the eigenmodes are:

\begin{equation}
  \psi\left(\vec x,\zeta_1\right)_{\vec k,n} = \exp \left(i \vec k \cdot \vec x +i n \frac{ \zeta_1 }{R}\right),
\end{equation}
and the associated eigenenergies are:

\begin{equation}
  E_{n}\left(\vec k\right) = \frac{\hbar^2}{2 m}\left( \vec k^2 + \frac{n^2}{R^2} \right).
\end{equation}

Here as well, there is a structure with continuous bands, with only one band appearing below an energy threshold of $\hbar^2/\left(2 m R^2\right)$.
Much like on the cylinder, there is no momentum on the circle at low energies, the usual relation between $3\text{D}$ momentum $\vec p$ and energy $E$ is realized: $E = {\vec p}^{\,2}/(2m)$, and the eigenfunctions are $3\text{D}$ plane waves $\psi_{\vec p} = \exp\left(i \vec p \cdot \vec x /\hbar \right)$\\

Once again, small extra dimensions are invisible to low-energy physics.

\subsection{Spherical extra dimensions}

In order to get a better grip of the general structure of extra-dimensional theories, let us now probe what would happen if instead of a circle, we consider two extra dimensions forming a sphere of radius $R$.
Here we will make the canonical choice for coordinates: $\zeta_1$ as the colatitude (often named $\theta$) and $\zeta_2$ as the longitude (often named $\phi$).
Wavefunctions will have to verify a more complex set of periodicity conditions, that reflect the structure of the sphere:

\begin{equation}
\begin{array}{c}
\psi(\vec x, \zeta_1, \zeta_2) =  \psi(\vec x, \zeta_1, \zeta_2+2 \pi ),\\
\psi(\vec x, \zeta_1+\pi, \zeta_2) =  \psi(\vec x, \pi - \zeta_1, \zeta_2+ \pi ).
\end{array}
\label{s2boundary}
\end{equation}

The generalization of the momentum square operator is $-\hbar^2 \Delta$, where $\Delta$ is the Laplacian, which is an operator that can be defined on any smooth curved space.

The free Hamiltonian on the sphere should therefore be

\begin{equation}
    {\cal H}_{S^2} =-\frac{\hbar^2}{2m}\left( \frac{1}{R^2} \partial_\theta^2 + \frac{1}{R^2\sin^2 \theta} \partial_\phi^2 \right).
    \label{s2lapl}
\end{equation}

The eigenfunctions for this operator are special functions called the spherical harmonics, which are constructed from the Legendre polynomials and automatically verify the periodicity conditions of Equation \ref{s2boundary}.
The spherical harmonics are denoted $Y_l^\mu$ and are indexed by two integers, $l\in \mathbb{N}$, and $\mu = -l \dots l$.\footnote{The most standard notation replaces $\mu$ by $m$ ($Y_l^m$), we however want to avoid any confusion with the particle mass, already named $m$.}

These function are typical special functions appearing in systems with spherical symmetry and have a wealth of well-known properties \cite{mathmeth}.
In particular, they are the eigenfunctions of the spherical Laplacian operator with eigenvalues $l(l+1)$, meaning that the eigenenergies are
\begin{equation}
  E_l = \frac{\hbar^2}{2m R^2}l(l+1),
\end{equation}
and are independent of $\mu$, leading to a $2l+1$ degeneracy for each level.

If we now upgrade to three usual dimensions and two spherical extra dimensions, the Hamiltonian becomes:

\begin{equation}
    {\cal H}_{3\text{D}\times S^2} =- \frac{\hbar^2}{2m}\Delta_{3\text{D}}- \frac{\hbar^2}{2m}\left( \frac{1}{R^2} \partial_\theta^2 + \frac{1}{R^2\sin^2 \theta} \partial_\phi^2 \right)
    \label{s2lapl}
\end{equation}
The eigenfunctions are
\begin{equation}
\psi\left(\vec x, \zeta_1, \zeta_2\right)_{\vec k,l,\mu} = \exp\left(i\vec k \cdot\vec x\right)Y_l^\mu\left(\zeta_1,\zeta_2\right),
\end{equation}
and the associated eigenenergies are

\begin{equation}
  E_l = \frac{\hbar^2}{2m}\left(\vec k^{\,2}+ \frac{l(l+1)}{R^2}\right).
\end{equation}

The lowest-lying mode is proportional $Y_{00}$, which is constant in $\theta$ and $\phi$ so the analysis of the previous cases repeats: below a threshold in energy of $\frac{\hbar^2}{mR^2}$, there is no sign of there being extra dimensions as the wavefunction is just  $\exp\left(i\vec k \cdot\vec x\right)$ and the energy is the usual $\vec p^{\,2}/2m$.

\subsection{General case}

Now that we have covered several examples, it is straightforward to generalize to arbitrary extra dimensions. The wave function will depend on $3+n$ coordinates $\left(\vec x, \zeta_i \right)$, where the $\zeta_i$ parametrize some space ${\cal M}$ whose Laplacian operator is $\Delta_{\cal M}$, which depends only on $\zeta_i$.

It is now easy to write the Hamiltonian:

\begin{equation}
{\cal H}_\text{3+n} = - \frac{\hbar^2}{2 m}\Delta_{3\text{D}} - \frac{\hbar^2}{2 m} \Delta_{\cal M}
\end{equation}

For a compact ${\cal M}$, the Laplace operator has a discrete set of eigenvalues $0<\Lambda_1<\Lambda_2<\dots<\infty$ and each eigenvalue $\Lambda_j$ has a finite number of associated eigenfunctions $\chi_{j}^{l}(\zeta_i)$, where $l=1,\dots,N_j$\cite{spectra}. In particular, it is always true that $\chi_{0}(\zeta_i)$ is a constant function.

A given space ${\cal M}$ can always be uniformly scaled up or down by multiplying all lengths by a given factor, which also scales the eigenvalues. It is a good idea to capture this behiavior with the scaling of the volume of ${\cal M}$, $\text{Vol}({\cal{M}})=L^n$: expressing the eigenvalues as $\Lambda_i = \lambda_i/L^2$ makes $\lambda_i$ independent of these global rescalings.

The eigenfunctions of our Hamiltonian are

\begin{equation}
  \psi\left(\vec x, \zeta_i\right)_{\vec k,j,l} = \exp\left(i \vec k\cdot \vec x\right) \chi_{j}^{l}(\zeta_i)
\end{equation}

and the associated eigenergies are:

\begin{equation}
  E_j\left(\vec k\right) = \frac{\hbar^2}{2m}\left(\vec k^{\,2}+ \frac{\lambda_j}{L^2}\right).
\end{equation}

We can see here that $L$ is the typical length scale associated with our space, as were the radii of the circle an sphere we considered in earlier sections, and scaling it up or down changes the energy at which extra-dimensional effects appear. For energies well below $\lambda_1\times \frac{\hbar^2}{2mL^2}$, or equivalently for wavelengths larger than $L/\sqrt{\lambda_1}$, the system we are describing is identical to a free 3-dimensional particle.

\section{Toward relativistic theories}
\label{relat}
As of Fall 2016 there is conclusive evidence that no particle excitation have been produced in collider experiments like the Large Hadron Collider below $\unit{1}{\tera\electronvolt}$\cite{uedlimit}, which is an energy an order of magnitude above the center of mass energy ($mc^2$) of the heaviest fundamental particle we know: the top quark ($m_{\text{top}}c^2=\unit{173}{\giga\electronvolt}$).

However, our non-relativistic derivation in Sections \ref{cyl} and \ref{3D} holds as long as the energy of the particle is below the relativistic rest energy $mc^2$.
In particular, if the excitation gap $\hbar^2/(2 m R)$ is comparable or higher than the rest mass, the spectrum we determined does not describe the system quantitatively above the first band.
This, however, is still enough to preserve the main conclusion of the previous sections: low energy physics is insensitive to the existence of small extra dimensions.
The global structure of the spectrum, with separate excitations that have a constant contribution on top of a continuous kinetic energy is preserved as well, but this constant extra-dimensional kinetic energy has new consequences once relativity is taken into account, which we shall now clarify.

While the proper way of doing relativistic quantum physics is quantum field theory, we will use the relativistic wave mechanics approach, which upgrades the Schr\"odinger equation to the Klein-Gordon equation\cite{KG}, as it captures enough of the complete picture to make interesting observations.
Much as the free Schr\"odinger equation is a quantization of the definition of the mechanical energy of a free non-relativistic particle $E=\vec p^{\,2}/2m$, the Klein-Gordon equation is a quantization of the relativistic definition of energy: $E^2=\vec p^{\,2} c^2 + m^2 c^4$ becomes
\begin{equation}
 \left( i\hbar \partial_t\right)^2 \psi = \left(\left(-i \hbar \vec\nabla \right)^2c^2 + m^2 c^4 \right)\psi.
\end{equation}

This equation defines the spectrum of the square of the Hamiltonian operator\footnote{Note that this means that negative energy solutions are possible.
Let us simply overlook them as a fully consistent relativistic quantum theory gives them an interpretation with positive energy}:
\begin{equation}
\hat{E}^2 \psi = {\cal H}^2 \psi = \left((\hbar c)^2 \Delta  + m^2 c^4 \right)\psi.
\end{equation}

In usual 3D space, this equation is not very different from the ones we already solved and it should now be clear that plane waves are again the eigenfunctions.
Inserting $\psi_{\Lambda} = \exp (i \vec k \cdot \vec x)$, we find that the eigenvalues of the square Hamiltonian are
\begin{equation}
\Lambda = E^2 = (\hbar c)^2 k^2  + m^2 c^4 = \vec p^{\,2} c^2 + m^2 c^4.
\end{equation}
Upgrading to a circular extra dimension is straightforward.
Diagonalizing the squared Hamiltonian
  \begin{equation}
  {\cal H}^2_{\mathbb{R}^3\times S^1} =  \left((\hbar c)^2 \left(\partial_x^2+\partial_y^2+\partial_z^2+\partial_\zeta^2\right) + m^2 c^4 \right)
  \end{equation}
  with plane waves, we find a quantized momentum in the extra dimension, yielding the same general structure of continuous bands:
\begin{equation}
E_n^2(\vec k) = (\hbar c)^2 \left(k^2+\frac{n^2}{R^2}\right) + m^2 c^4.
\end{equation}

As in the non-relativistic approach, there is a gap below which the energy-momentum relation is identical to the case without extra dimensions: $E^2 = \vec p^{\,2} c^2 + m^2 c^4$.
The first excitation appears for energies above $(\hbar c)/R$, which is now independent of the mass, but conserves the feature that smaller dimensions require larger energy to be probed.

An interesting consideration is the behavior of the excitation from the point of view of an observer in 3D. It is clear that, for a $3+n$D observer, it is just a free particle moving in $3+n$D space.
A 3D observer, however, would see an excitation as a particle moving with an arbitrary momentum $\hbar \vec k$ in 3D space.
A general method for measuring the mass of particles\cite{mmeas} is to measure their energy and their momentum independently and compute
\begin{equation}
  m^2_{\text{obs}} = \frac{1}{c^4}(E^2-\vec p^{\,2} c^2).
\end{equation}
For particles in the first excited level, one would therefore find
\begin{equation}
m_{\text{obs}}^2 = m^2 + \left(\frac{\hbar}{R c}\right)^2,
\end{equation}
the excitations behave like a heavier copy of the low-energy particle.
These heavy copies are known in the literature as Kaluza-Klein excitations \cite{kk}.
This has a simple explanation: the kinetic energy in the extra dimension is not seen by a 3-dimensional observer and therefore constitutes a contribution to the energy in the rest frame of the particle.

Kaluza-Klein particles are how particle collider experiments can probe the possibility of extra-dimensions. A complete model with interactions between particles would predict the probability for particle collisions to produce excitations in collisions, and the non-observation of new heavy particles allows limits to be set on the size of possible extra-dimensions. Since Kaluza-Klein particles with a mass below $\unit{1}{\tera\electronvolt}$\cite{uedlimit} have been excluded, we know that if extra-dimensions exist, they have to be of a radius $\unit{R< 2\times 10^{-19}}{m}$.

\section{Conclusion}
Physics students who take a first quantum mechanics course are often eager to learn with more precision the ideas that are presented in popular science, which therefore seems to do a good job of engaging younger people by teasing them with the unexpected developments of modern physics. I believe that it is important to feed this enthusiasm by making contact, when convenient, between the classroom and this now large body of popularized materials. While it is difficult to provide a serious introduction to extra-dimensional theories or to string theory at the undergraduate level, this paper attempts to extract two features of extra dimensions and present them in a format that is both short and elementary. The advanced reader who would want to go further is encouraged to read graduate-level lectures and research-level reviews as for example Ref.~\onlinecite{tasi} and Ref.~\onlinecite{edpdg}. An interesting collection of references is also found in Ref.~\onlinecite{backre}.

This paper is intended for university educators as well as students and is therefore accompanied with an example of an exercise which can be proposed as a follow up of the particle in a box. A translated version has been submitted to a class of French $3^{\text{rd}}$-year undergraduate students at the \'Ecole Sup\'erieure de Lyon who were at the middle of their first quantum mechanics course. Most students having been able to solve it, it should be expected that this exercise is well suited to undergraduate courses. I would be happy to hear about other attempts at confronting this exercise or variations to students.

\section*{Acknowledgements}
This work was partly developped at the Korean Institute of Advanced Studies in Seoul, the Center for Theoretical Physics of the Universe in Daejeon, whose hospitality is greatly appreciated. I would like to thank Aldo Deandrea for his suggestions on the manuscript and his support and Baptiste Demoulin for his useful comments on the exercise and for proposing it to his class at the ENS Lyon.

\clearpage

\appendix

\begin{widetext}
\section*{Appendix: From the particle in a box to theories with compact extra-dimensions}
\label{exo}
\textit{This is intended as an application of the classical exercise of a quantum particle in a box.
The students should have solved the Schr\"odinger equation and have the spectrum, either with hard wall boundary conditions or with periodic boundary conditions}
\subsection{Free particle on a cylinder}
Let us consider a free particle in two dimensions with the usual Hamiltonian operator ${\cal H} = - \hbar^2/(2m)\left(\partial_{x_1}^2 + \partial_{x_2}^2\right)$.
We will model a free particle on a cylinder with radius $R$ by imposing boundary conditions on the wavefunction.
\begin{enumerate}
  \item If $x_1$ is the direction \textit{along} the cylinder and $x_2$ is the direction \textit{around} the cylinder, what boundary condition should the wavefunction $\psi(x_1,x_2)$ obey?
  \item Based on your experience with free particles and free particles in a box, propose an Ansatz for the eigenfunctions of the Hamiltonian.
  \item Using your Ansatz, give the set of eigenfunctions of ${\cal H}$ and the associated eigenvalues.
  \item The eigenvalues are given by a continuous parameter and a discrete parameter.
Draw a sketch of the eigenenergies as a function of the continuous parameter for the first lowest values of the discrete parameter.
  \item Show that below a certain energy, the free particle on a cylinder has the same behavior as a simple one-dimensional system.
\end{enumerate}

\subsection{Extra dimensions}
In string theory, the usual $3+1$ dimensional spacetime is augmented by 6 other compact directions.
As a simpler illustration of the consequences of this possibility, let us study the addition one single compact dimension to the 4 usual ones.
In practice, we will study the spectrum of a free particle with eigenfunction $\psi(x,y,z,\zeta)$ where $\zeta$ obeys the same boundary condition as $x_2$ in the previous section.

\begin{enumerate}
  \item Express the new Hamiltonian in this situation and find its spectrum using the results of the previous section.
  \item Show that at low energies, this particle behaves like a well known 3D system.
\end{enumerate}

\textit{This part requires some knowledge of special relativity}

While we are looking at a nonrelativistic system, it is interesting to think about the relativistic generalization of our conclusions.
The structure of continuous bands separated by jumps of a discrete quantum number is still present, but the spectrum is now $E_n^2(\vec p) = \vec p^{\,2} c^2 + (\hbar c)^2 \frac{n^2}{R^2} + m^2 c^4$ where $n\in \mathbb{N}$ and $p$ is the 3-momentum.

\begin{enumerate}
  \item Propose a method for obtaining the mass of a particle based on measurements of its kinematic properties
  \item If one observed such a particle with level $n=1$ and momentum $\vec p$, what result would this mass measurement produce?
\end{enumerate}

\end{widetext}

\end{document}